\documentclass[epj,final]{svjour}
\usepackage{epsfig}
\usepackage{amsmath,amssymb}
\begin{document}
\newcommand{\vm}{v_{\text{max}}}
\newcommand{\vs}{v_{\text{safe}}}
\newcommand{\vst}{\tilde{v}_{\text{safe}}}
\newcommand{\vd}{v_{\text{des}}}
\newcommand{\vdt}{\tilde{v}_{\text{des}}}
\newcommand{\va}{v_{\text{anti}}}
\newcommand{\lc}{l_{\text{car}}}
\def\eqref#1{equation (\ref{#1})}
\def\Eqref#1{Equation (\ref{#1})}
\def\figref#1{figure~\ref{#1}}
\def\Figref#1{Figure~\ref{#1}}
\def\secref#1{section~\ref{#1}}
\newcommand{\SK}{SKM}
\newcommand{\NSK}{SKA}
\title{\bf Effects of anticipatory driving in a traffic flow model}
\author{Nils Eissfeldt\inst{1} \and Peter Wagner\inst{2}}
\institute{ZAIK -- Center for Applied Informatics, Universit\"at zu
  K\"oln, 50931 K\"oln, Germany
  \and
  Institute for Transportation Research, German Aerospace Center, 12489
  Berlin, Germany
  }
\authorrunning{N.~Eissfeldt and P.~Wagner}
\titlerunning{Effects of anticipatory driving in a traffic flow model}
\date{}
\abstract{Anticipation in traffic means that drivers
  estimate their leaders' velocities for future timesteps. In the
  article a specific stochastic car--following model with non--unique
  flow--density relation is investigated with respect to anticipatory 
  driving. It is realized by next--nearest--neighbour interaction
  which leads to large flows and short temporal headways. The
  underlying mechanism that causes these effects is explained 
  by the headways of the cars which organize in an
  alternating structure with   a short headway following a long one,
  thereby producing a strong  anti-correlation in the gaps or in the
  headways of subsequent cars. For the investigated model the
  corresponding time headway distributions display the short headways
  observed in reality. Even though these effects are discussed for a
  specific model, the mechanism described is in general present in
  any traffic flow models that work with anticipation.} 
\PACS{{02.50.Ey}{Stochastic processes} -- {45.70.Vn}{\ Granular models
    of complex systems; traffic flow} -- {89.75.Fb}{\ Structures and
    organization in complex systems}} 
\maketitle
\section{Introduction}
The basic mechanisms that are responsible for traffic flow breakdown
are still not very well understood and discussed controversial
\cite{ker98,dag99,cho00,hel01}. One reason for it is, that the
microscopic models in use nowadays still have deficiencies, however it is not 
obvious which ones. Therefore it is not clear whether the mechanisms
of breakdown displayed by a certain model have a counterpart in
reality. 

Sometimes, even the inner working of those models is not very well
understood. This is true, e.g., for the models that work with
so-called anticipation \cite{kno00,jia03}. Here, anticipation means
that drivers estimate the velocity of preceding cars for future time
steps. With respect to safe car motion this driving strategy avoids
abrupt braking and therefore leads to a stabilization of the flow in
dense traffic \cite{kno00}. As a result, these models display
small temporal headways that are similar to the ones observed in
reality. Although the mechanism of stabilization of the flow
seems to be necessary with respect to the reproduction of real--world
traffic data \cite{unpub}, the changes in the model's dynamics in
consequence of anticipatory driving are not known in detail.
Therefore, it is the aim of this  article to clarify the role of
anticipation in microscopic traffic flow models.

This will be done for a certain well--known microscopic traffic flow
model which is described in \secref{sec_SK_Model}. In this model, 
anticipation is introduced via next--nearest--neigh\-bour
interaction. The consequences on the model's dynamics is explored by
simulation as well as analytical calculations
(cf. \secref{sec_erg}). Leaping ahead, it is stated that by virtue  
of anticipation the system organizes the headways of the cars in an
alternating structure which allows for the observed small temporal
headways. As will become clear from the discussion, most of the results 
found in the following should be at work in other models, too.
\section{A car--following model with anticipation}
\label{sec_SK_Model}
As stated above, the effects of anticipation will be investigated
using a specific car following--model. The model described in
\cite{kra97,kra98,kra98b} is used as reference model and is
referred to as \SK\ in the following.

It is based on an approach by Gipps \cite{gip81} and three basic 
assumptions, namely
\begin{itemize}
\item that vehicles move collision--free,
\item not faster than a maximum velocity $\vm$ and
\item individual car acceleration $a$ and deceleration $b$ are
  bounded. 
\end{itemize}
Based on the requirement of collision--freeness a safety--condition
can be derived. Assume one car (driver--vehicle unit)
with velocity $v$ is following another car
(driving with velocity $\tilde{v}$) within a distance $g$. Here, $g$
is the free space between vehicles, i.e., the distance between the
cars at positions $x$, $\tilde{x}$ minus the cars' length
$\lc$. Safety, i.e., crash--free motion is guaranteed if
\begin{eqnarray}
  \label{eq:safety_condition}
  d(v) + \tau v \leq d(\tilde{v}) + g
\end{eqnarray}
holds, with $d(v)$ being the braking distance needed to stop when
driving with velocity $v$ and $\tau$ a finite reaction time.
For braking with constant deceleration $b>0$, i.e., $-b \leq dv / dt$
the braking distance is given by $d(v) = v^2 /( 2b)$. The
\eqref{eq:safety_condition} then leads to
\begin{eqnarray}
  \label{eq:vsafe_sk}
  \vs (\tilde{v},g) = -b\tau + \sqrt{b^2\tau^2 + \tilde{v}^2 + 2bg}.
\end{eqnarray}
In order to complete the definition of the model's dynamics it is
assumed that every car moves at the highest velocity compatible with the
assumptions. Based on these assumptions an update scheme can be
formulated in the manner of the well--known Nagel--Schreckenberg model 
(NaSch) \cite{nag92}.
\begin{figure}[!t]
  \centerline{\epsfig{figure=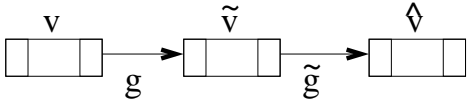,width=0.7\linewidth}}
  \caption{Visualization of the variables $v, g, \tilde v, \tilde
    g, \hat{v}$. All cars are considered to have equal length $\lc$.}
  \label{fig:params}
\end{figure}

The \SK\ is defined with continuous state variables $x,v$ and
discrete timesteps $\Delta t$. In each timestep every car is updated
after calculating its $\vs$ according to the following scheme
\begin{eqnarray}
  \nonumber
  \vd &=& \min \{v_t + a \Delta t,\vs,\vm \}\\
  \label{eq:scheme_sk}
  v_{t+\Delta t} &=& \max \{\vd - \eta \epsilon a,0
  \}\\  
  \nonumber
  x_{t+\Delta t} &=& x_t + v_{t+\Delta t} \Delta t.
\end{eqnarray}
The update (\ref{eq:scheme_sk}) is done in parallel. The random
fluctuation of strength $\eta \epsilon a$ is introduced to mimic
deviations from the optimal driving strategy given by $\vs$. $\eta$ is
a random number uniformly distributed in the interval $[0,1]$ and the
parameter $\epsilon$ determines the fluctuation strength in units
of $a$.

Before introducing anticipation into the model it should be
stated that the formulation of $\vs$ in \eqref{eq:vsafe_sk} differs
from that given in \cite{kra98}. The reason is mostly that the
calculation of the update scheme (\ref{eq:scheme_sk}) becomes more
easy. However, due to its structure we could not
find a proof for crash-freeness analytically (as is possible for the
original formulation). But extensive simulations with $\tau \geq
\Delta t$~\footnote{This conditions simply states that safe driving is
  possible, if the ``true'' reaction time, i.e., one timestep, is
  smaller or equal to the reaction time each driver assumes.}  neither
gave a hint for collisions nor had we found a crucial difference in
the model's dynamics.

We recall that in the \SK\ each car only takes into account the
car in front to deduce its optimal driving strategy. It is common experience that such assumption is unrealistic, especially in
dense traffic situations. In order to bring anticipation into the
model the update scheme is modified by an intermediate step: Each
driver predicts the worst--case strategy $\va$ her predecessor will
choose in the next timestep. Assuming that there is a car in front of
the predecessor within a distance $\tilde{g}$ driving with velocity
$\hat{v}$ (see \figref{fig:params}), then
\begin{eqnarray}
  \label{eq:vanti}
  \va = \max \{ \vdt - \epsilon a,0\}
\end{eqnarray}
with
\begin{eqnarray}
  \label{eq:vdes_tilde}
  \vdt = \min \{ \tilde{v} + a, \vst(\hat{v},\tilde{g}), \vm \}. 
\end{eqnarray}
The calculated $\va$ will then be used to determine the safe velocity.
Therefore, the safety condition \eqref{eq:safety_condition} is
restated with the assumption that the leading car will choose
$\tilde{v}_{t+\Delta t} \geq \va$ as driving strategy,
\begin{eqnarray}
  \label{eq:safety_restated}
  d(v) + \tau v + \gamma_c(v,\tilde{v}) \leq d(\va) + \va \tau + g.
\end{eqnarray}
The function $\gamma_c(v,\tilde{v})$ has been introduced to take into
account ``unexpected'' fluctuations in the predecessor's driving
behaviour. Then, the new expression \eqref{eq:safety_restated} leads
to a new expression for the safe velocity,
\begin{eqnarray}
  \nonumber
  \vs = &-&b \tau \\
  \label{eq:new_vsafe}
  &+& \sqrt{b^2\tau^2 + \va^2 + 2b(g+\va\tau-\gamma_c(v,\tilde{v}))}.
\end{eqnarray}
In the following,
\begin{eqnarray}
  \label{eq:gc}
  \gamma_c(v,\tilde{v}) = \min \{\va\,\tau,g_c\},
\end{eqnarray}
will be chosen where $g_c$ is constant. Since
$g+\va-\gamma_c(v,\tilde{v})$ can be interpreted as an effective gap
$g_{\text{eff}}$, where \eqref{eq:gc} forces $g_{\text{eff}} \geq g$.
The idea of the effective gap is similar to the
cellular--automaton model in \cite{kno00} (BL--CA).
The major difference is that in the modified \SK\ anticipation
enters into the model by velocity and the effective gap
(cf. \eqref{eq:new_vsafe}) while in the
BL--CA it does just via the latter.

Besides the new definition of $\vs$ the update scheme
(\ref{eq:scheme_sk}) is used.
\section{The role of anticipation}
\label{sec_erg}
\begin{figure}[!t] 
  \centerline{\epsfig{figure=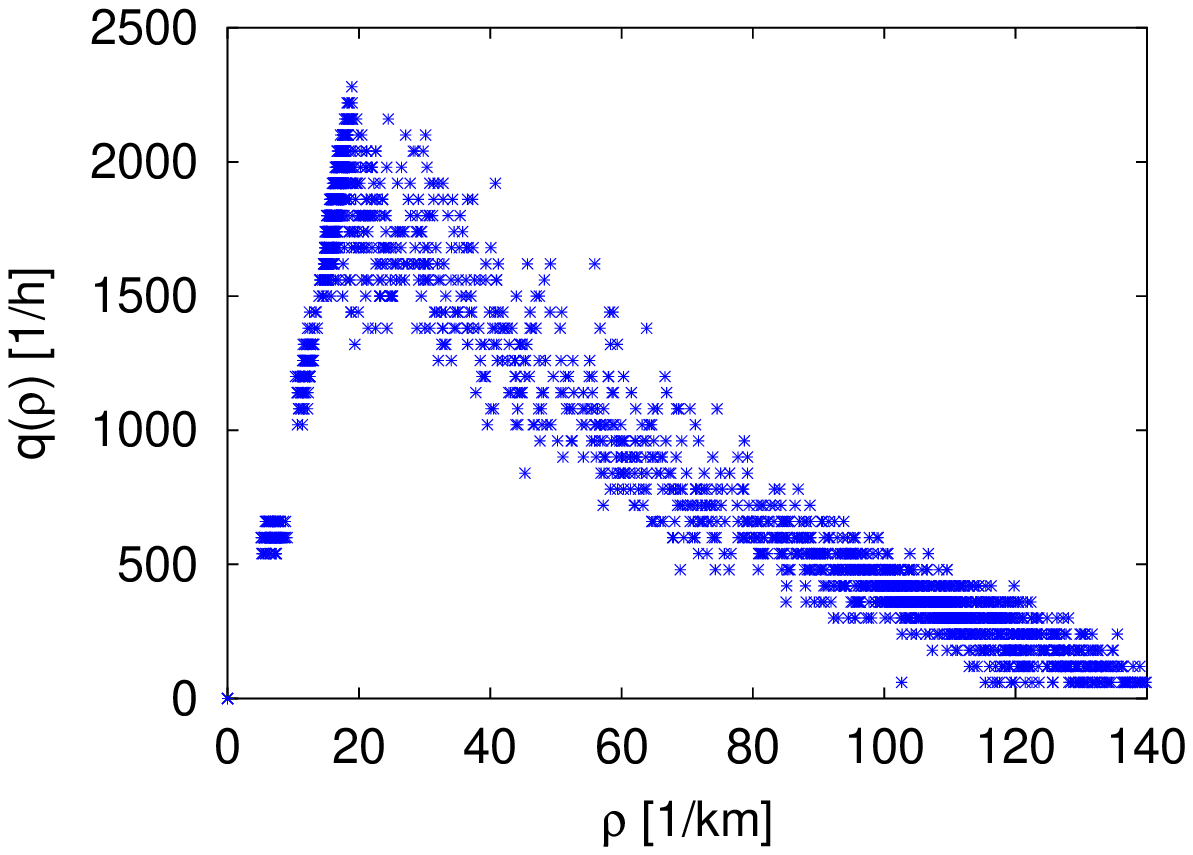,width=0.7\linewidth}}
  \centerline{\epsfig{figure=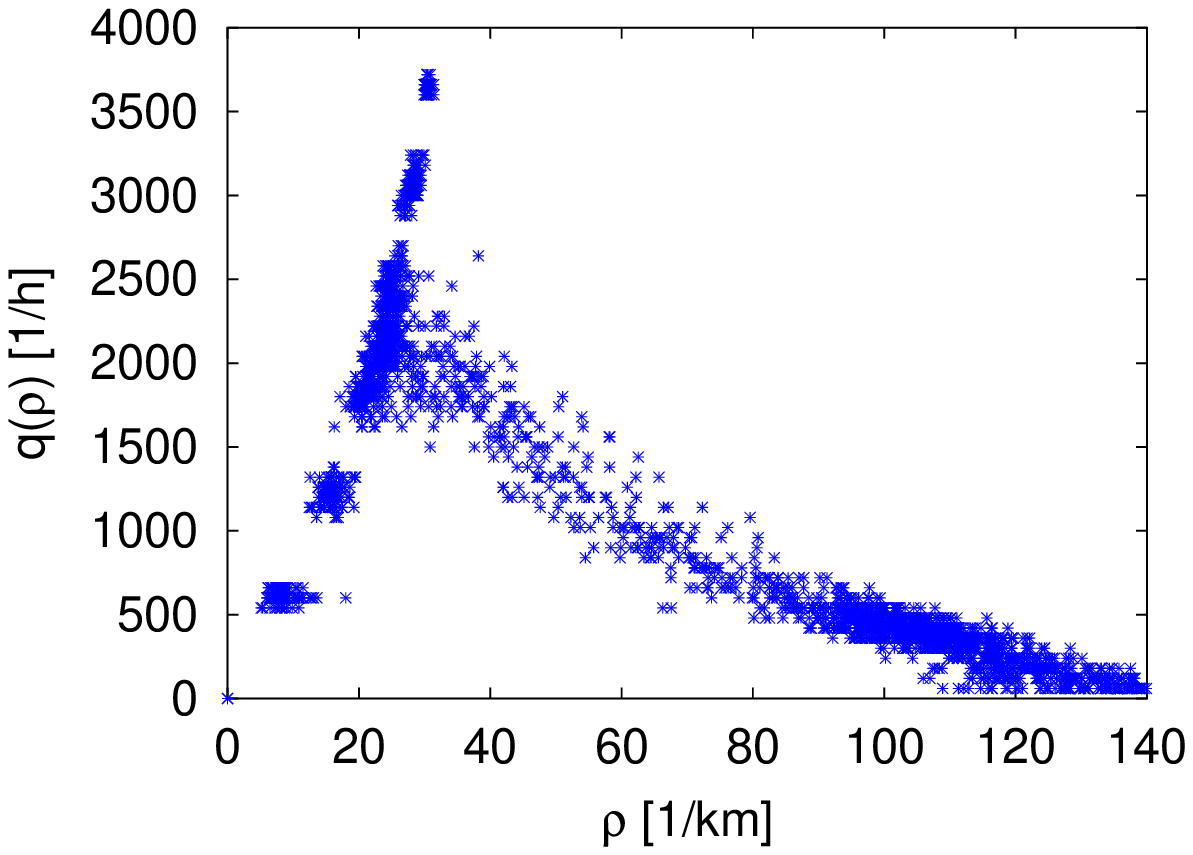,width=0.7\linewidth}}
  \centerline{\epsfig{figure=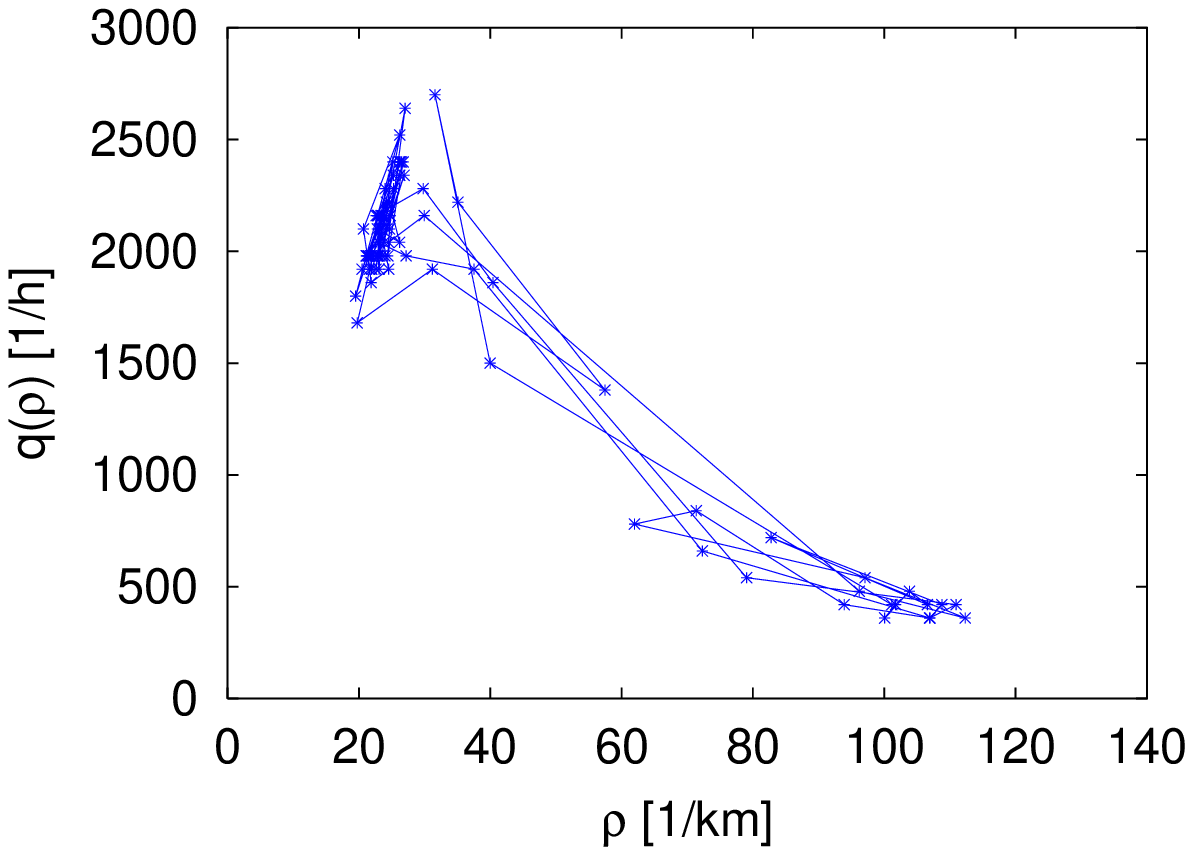,width=0.7\linewidth}}
  \caption{Plotted is the ``local'' fundamental diagram for the 
    \SK\ (top), \NSK\ (middle) as function of the density, and for the \NSK\
    where the global density has been fixed to $\rho = 35 km^{-1}$
    (bottom), a
    region where jammed and free flow coexist. Local means that the
    plot is obtained by mimicking a loop detector, and the
    time-averaging is done only over 60 s.}
  \label{fig:3nsk_fdg}
\end{figure}
In this section, by means of computer simulations, the \SK\ with
anticipation (\NSK) will be compared to the original model. For this
purpose, a fixed set of parameters is used, namely
\begin{eqnarray*}
  \begin{array}[c]{lp{0.5cm}lp{0.5cm}l}
    a = 2 \; m/s^2 && b~=~8~\;~m/s^2 && \vm~=~35~\;m/s\\
    \epsilon~=~1 && g_c=1 && \lc~=~7\;~m
  \end{array}
\end{eqnarray*}
As time scale $\tau = \Delta t = 1~\;s$ is chosen.
With respect to these parameters jam formation (wide moving jams) and
stable high-flow states exist in the corresponding \SK\ 
\cite{kra98,nag02} (cf. following subsection).
\subsection*{Flow--density relation}
To get started both models were simulated using periodic boundary
conditions, i.e., on a one-lane loop. In order to measure the
flow-density relation the loop was initialized homogeneously at
different global densities. After relaxation of the system mean
density $\langle \rho \rangle$, mean velocity $\langle v \rangle$ and
flow $\langle q \rangle$ were measured at a fixed location using
$60\;s$ intervals for sampling. The local density for a car $n$
passing the counting location is defined as
\begin{eqnarray}
  \rho_n = 1 / ( g_n + \lc )
\end{eqnarray}
Comparing the flow-density relations of the models
(\figref{fig:3nsk_fdg}) they both display a high-flow state and a
capacity drop at intermediate densities. The latter indicates
slow-to-start behaviour. Note that there is no explicit rule
introducing this effect and it results from
the the asymmetry in the randomization process for small speeds.
As can be seen, this mechanism is not changed by the introduction of
next-nearest-neighbour interactions. Moreover, the "optimized" driving
strategy even leads to a stabilization of the high-flow branch towards
higher densities compared to the \SK\ as already stated.

\begin{figure}[!t]
  \centerline{\epsfig{figure=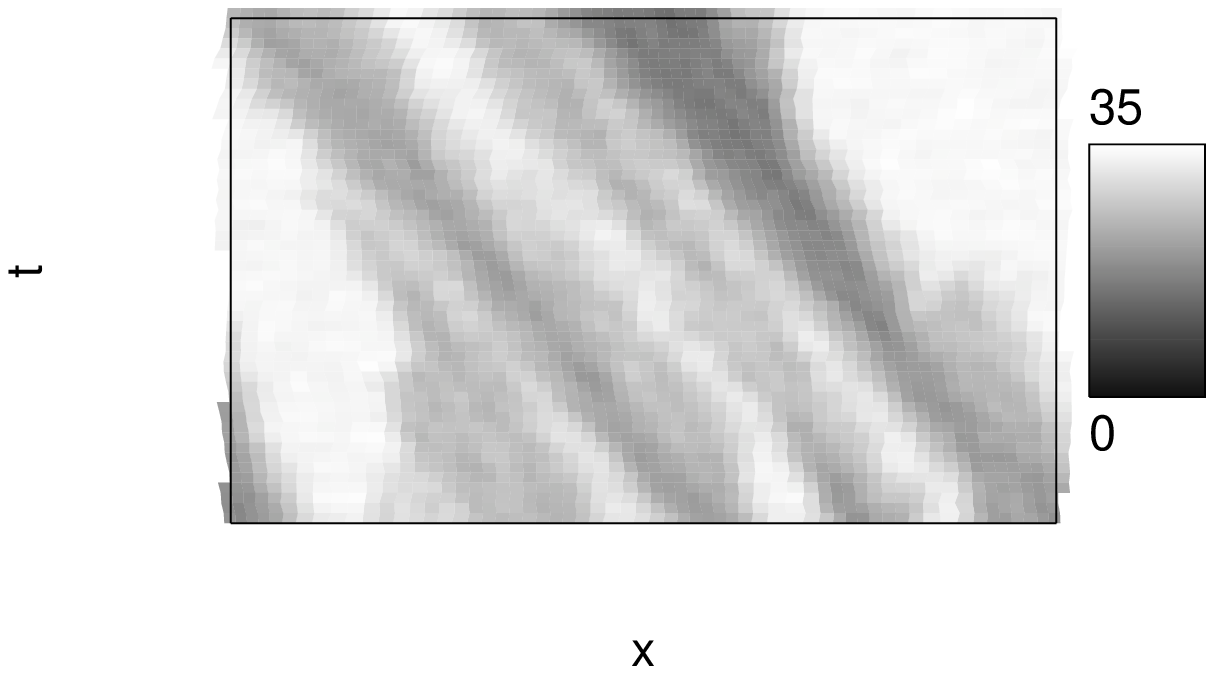,width=0.7\linewidth}}
  \centerline{\epsfig{figure=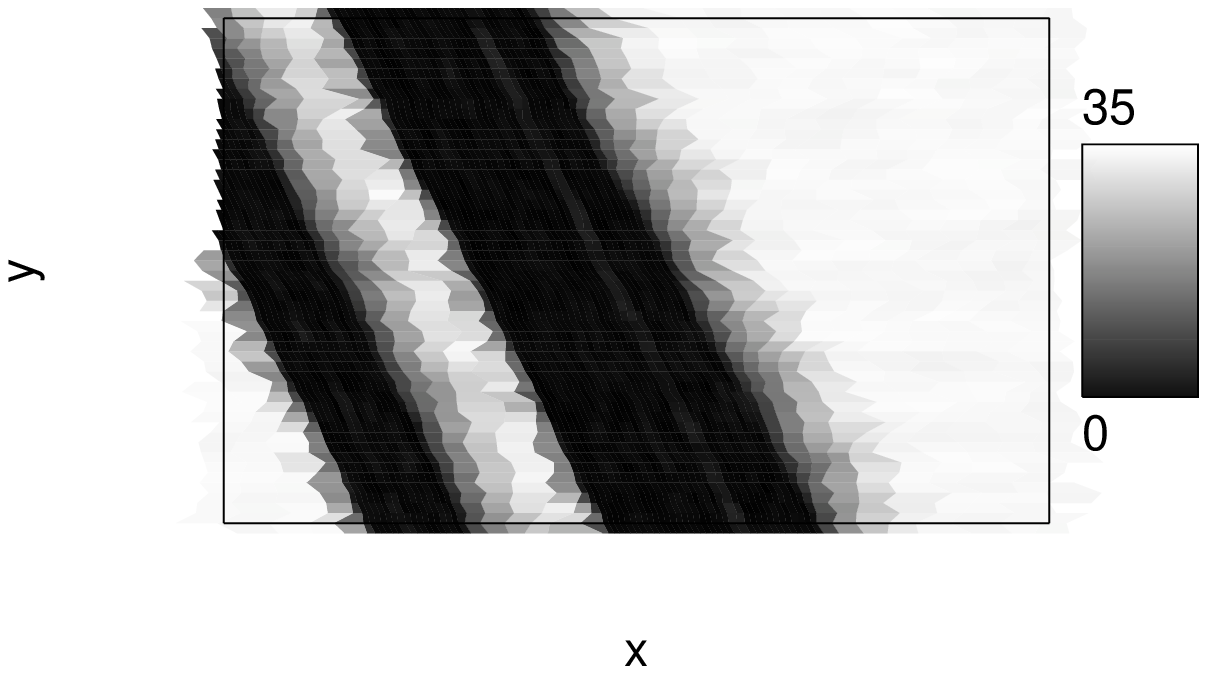,width=0.7\linewidth}}
  \caption{Space-time-diagram for a typical evolution of the
    \NSK. Each car is coloured by its current velocity in $m/s$.
    Initially, there are a lot of small jams (top) 
    that coagulate into a
    double jam state (bottom). 
    However, when waiting for a very long time, only 
    one wide moving jam remains. }
  \label{fig:3nsk_xt}
\end{figure}

In the closed system the jam state co--exists with the free--flow
state for densities $\rho \geq 20 km^{-1}$. Time-series at a fixed
density in that regime therefore display free-flow and jammed states
alternately (\figref{fig:3nsk_fdg}). At densities where the
homogeneous free-flow state is unstable, small clusters of cars are
generated due to intrinsic fluctuations. In the long run, this number
of jams decreases until only one jam is left (\figref{fig:3nsk_xt}).
Recalling the arguments of \cite{kra98} the jamming transition in the
\NSK\ is a phase transition and one finds a
phase-separated system at equilibrium using periodic boundary
conditions. 

\begin{figure}[!t]
  \centerline{\epsfig{figure=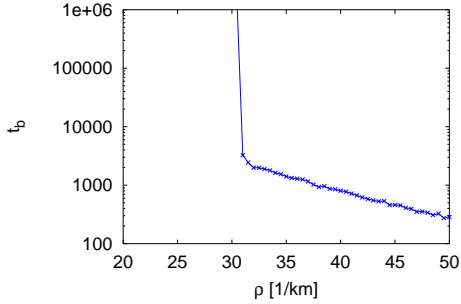,width=0.7\linewidth}}
  \caption{The average break-down time $t_{\text{b}}$
    in seconds for the \NSK. 
    For any density, a system with 5000 cars has been initialized with
    the homogeneous state. After waiting for at most $10^6$ time-steps
    and repeating this for 20 different realizations, the average time
    to breakdown can be approximated.}
  \label{fig:3nsk_tbrake}
\end{figure}

In \cite{nag02} a classification for stochastic traffic flow models is
provided based on the breakdown mechanism. To be more precise, in the
density regime where the jam solution co--exists with the high--flow
state one distinguishes two classes. Models are said to have a
``stable outflow'' if intrinsic fluctuations are not able to trigger
the transition from homogeneous flow to the congested phase. With
respect to the chosen parameters, the \SK\ belongs to that class of
models. In contrast, models like the VDR-model \cite{bar98} display
real metastability in that density regime and are said to have
``unstable outflow''. In \figref{fig:3nsk_tbrake} the waiting time
until the first stopped car is found is 
shown versus the
system's density. For each run a system with 5000 cars was initialised 
homogeneously. The values presented are means of 20 realisation per
density. As can be seen, at $\rho_c \approx 31 km^{-1}$ this time
diverges, i.e., homogeneous states corresponding to $\rho \leq \rho_c$
are stable. These state correspond to the high-flow branch in
the flow-density relation (\figref{fig:3nsk_fdg}) and therefore , the
\NSK\ owns the same type of bistability \cite{nag02} as the \SK\
does. 

By virtue of the anticipatory driving strategy, the maximum attainable
flow in the \NSK\ is unrealistically high, which is a known
feature for some extensions of the NaSch--model as well\cite{unpub}.
In that case the 
high flows only occur, if anticipatory driving as
defined in \cite{kno00} is switched on in, e.g. the VDR--model
\cite{bar98}, without changing the model's parameters.

Even though such states only appear for special initial conditions,
i.e., highly ordered homogeneous configurations, modifications are
necessary to use it in reasonable applications. However, it is a
different question whether these flows can be attained in a realistic
settings with macroscopic disturbances from lane-changing, on- and
off-ramps etc. Additionally, by introducing a diversified driver
behaviour also might lower the maximum flow to realistic values. For
example this might be done is by using a distribution $p(\tau)$
for the parameter $\tau$ so that each driver has her individual
$\tau_i$ drawn from that distribution. Another way of doing it
is to increase
$g_c$. Nevertheless, we will not take into account such problem
since we concentrate 
at discussing the effects of anticipatory driving on
the \SK.

Apart from these unrealistic high flows, it can be concluded that the
all--over macroscopic properties of the \NSK\ under periodic boundary
conditions (i.e., the global fundamental diagram, the spontaneous jam
formation or the existence of compact jams) are similar to the
corresponding \SK.
\subsection*{Time--headway distribution}
Several empirical studies have analyzed single vehicle data from
counting loops \cite{kos83,ban94,neu99,kno02,til00}. Such measurements 
provide in\-for\-ma\-tion about the microscopic structure of traffic\\
streams. The investigation of the corresponding observables in
stochastic traffic flow models can therefore justify their quality.  
\begin{figure}[!t] 
  \centerline{\epsfig{figure=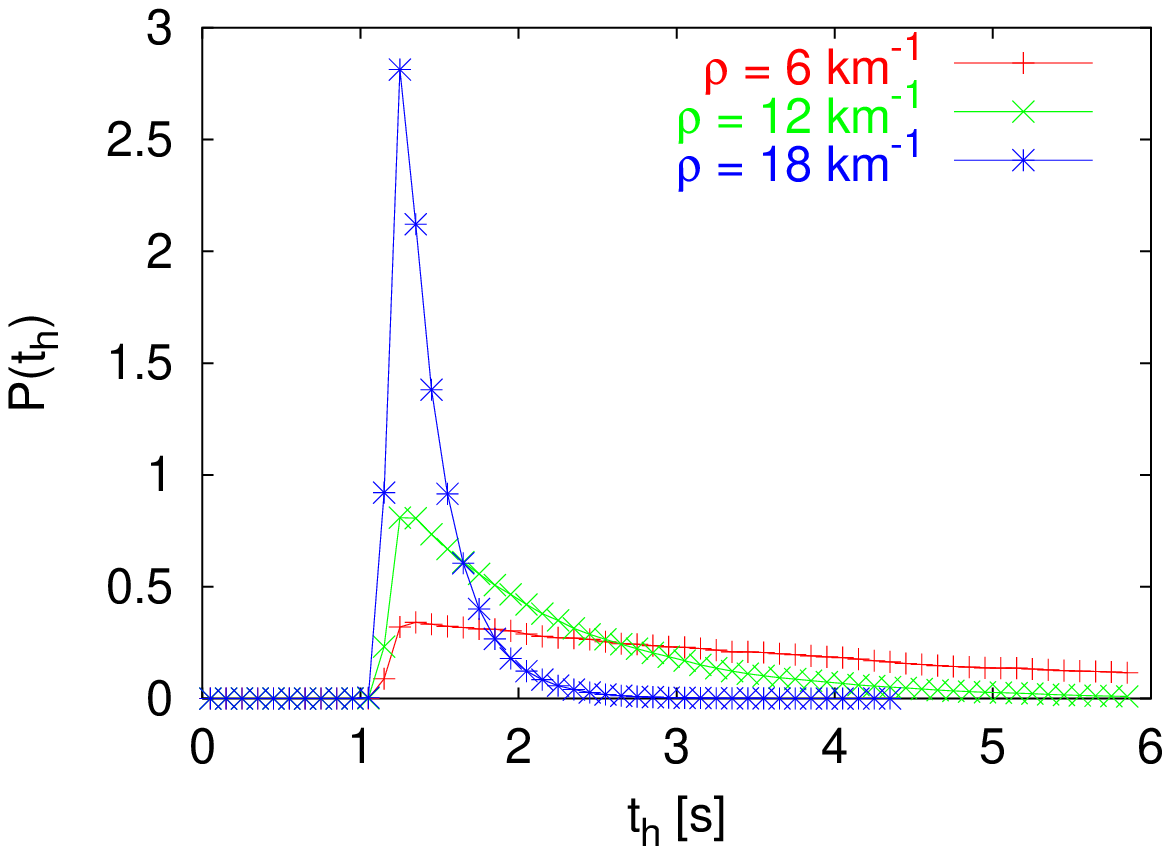,width=0.7\linewidth}}
  \centerline{\epsfig{figure=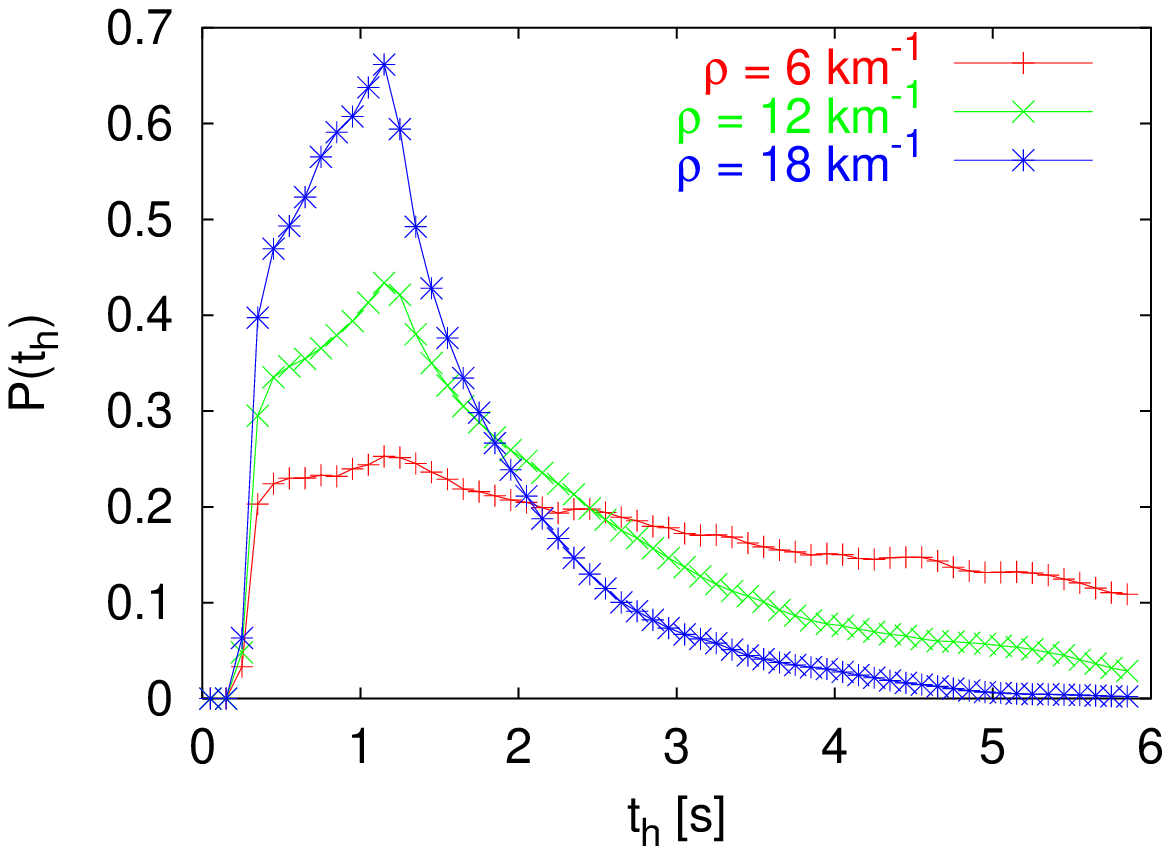,width=0.7\linewidth}}
  \caption{Time headway distribution for the \SK\ (top) and \NSK\ (bottom) 
    under free-flow conditions.}
  \label{fig:th_free}
\end{figure}

The time-headway is the microscopic analogue of the inverse flow. In
real data, it is simply measured by the time-difference $t_h = t_{i-1}
- t_i$ between the times of two cars passing the observer. Since this
model has a time-step of $\Delta t = 1\;s$, a different approach has
to be used to measure the time headway between two cars. This is 
done by using the relation
\begin{eqnarray*}
  t_h = g / v.
\end{eqnarray*}
The closed loop still serves as the computer-experimental setup. It is
initialized at different densities and the time-headway distribution
is measured after a sufficient relaxation time. 

In \figref{fig:th_free} the time-headway distribution of the free
flow phase at different densities is presented. From empirical
investigation it is known that  
in free flow extremely small time-headways exist ($t_h^{min} \approx
0.2 \; s$). Moreover, the maximum of the distribution and its shape at 
short times are independent of the density \cite{neu99,kno02,til00}.

\Figref{fig:th_free} (top) shows that the original \SK\ is not able to
reproduce such small time-headways in the free flow phase. There
exist a sharp cut-off at $t_h~\approx~\Delta t~=~1\;s$, i.e., 
the model's dynamics leads to $v_{t + \Delta t} \leq g_t$.
The maximum of the distributions is
located at $t_h \approx 1.3\;s$. Since in free flow $\langle v \rangle
\approx 34\;m/s$ this corresponds to $\langle g \rangle \approx
41.5\;m$, i.e., the \SK\ owns a fix point in its dynamics (for more
details see next subsection)

In contrast, the time-headway distribution of the \NSK\ shows a
broader peak structure (\figref{fig:th_free} (bottom)) and headways
noticeable smaller than $1 \; s$ exist, just as in empirical
observations. Even more, the distribution at short times is
independent of the density. 
However, the broadness of the peaks is not found in real--world
observations.
The occurrence of short time-headways stems
from the introduction of the velocity anticipation. Drivers can
optimize their gap to the leading vehicle since they have an idea 
about its future be\-haviour.
Smal\-ler gaps at $v \approx \vm$ than in the \SK\ are therefore
allowed.

Moreover, the broadness of the peaks indicates that in the \NSK\ a
range of gaps can be taken by cars driving at $v \approx \vm$ in the
free flow phase, i.e., there is no such strong fixed--point in the
car-following dynamics as is in the \SK\ (leaping ahead, this results 
from the fact that in the \NSK\ two consecutive cars share their
common gap $g + \tilde{g}$). With increasing density the peak in the
distribution becomes more and more pronounced and is shifted towards
smaller time-headways. The position of the peak correspond to the mean
gap, given by the initial conditions, $\langle g
\rangle~=~1~/~\rho~-~\lc$. Moving towards higher densities along the
free-flow branch of the flow--density relation the possible range of
gaps between cars decreases.

\begin{figure}[!t] 
  \centerline{\epsfig{figure=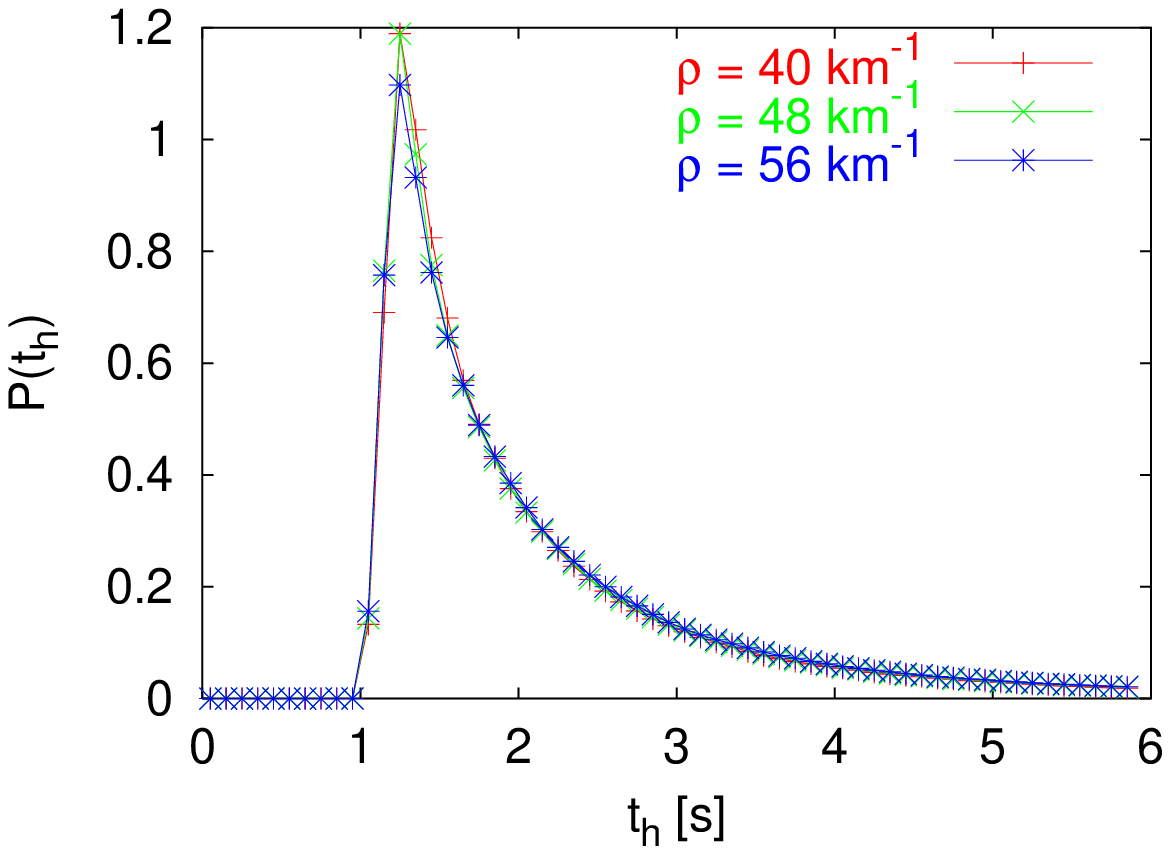,width=0.7\linewidth}}
  \centerline{\epsfig{figure=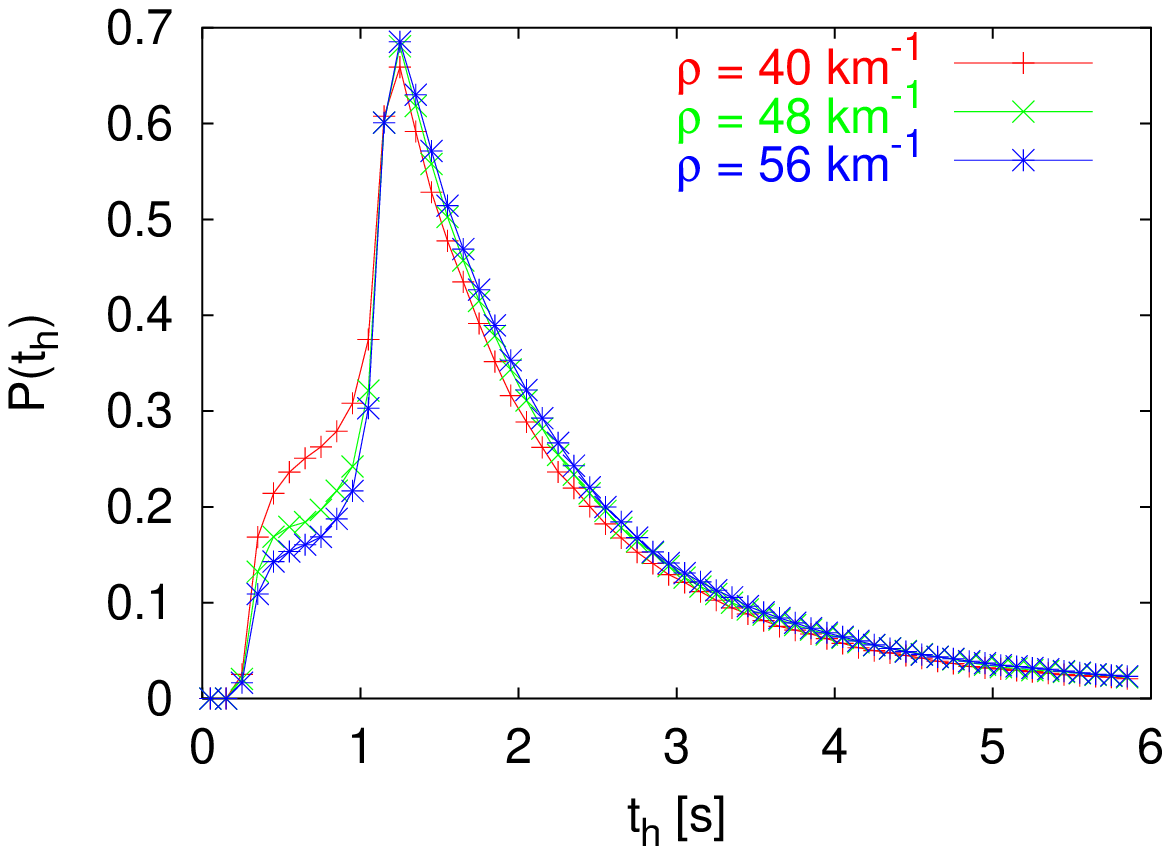,width=0.7\linewidth}}
  \caption{Time headway distributions for the \SK\ (top) and \NSK\ (bottom) 
    in congested flow. Headways smaller than $1\;s$ stem from the free
    flow state, their share decreases with increasing density.}
  \label{fig:th_jam}
\end{figure}

Comparing the time-headway distributions in the congested state
(\figref{fig:th_jam}), they are for both models almost independent of
the density. The exponential decay of the distribution results from
the fact that for large headways cars can be regarded as almost
independent from each other, implying a Poissonian distribution.
They have their maximum around $t_h \approx 1\;s$ which agrees with
empirical findings. Unlike reality the peak is fixed and not as broad.

From this we conclude that concerning the dynamics inside jams both
models behave similarly.

The appearance of $t_h \leq 1$ in the case of the \NSK\ is due to the
experimental setup. In the closed loop the system is separated into
two phases, one wide moving jam and a region of free-flow
(\figref{fig:3nsk_xt}). As demonstrated before, time--headways smaller
than $1\;s$ can be found in the free--flow phase. Therefore, cars that
are not in the jammed state generate these time-headways. Since the
number of cars in the free phase decreases with increasing density the
weight of small time--headways also reduces.
\subsection*{Optimal velocity curve}
Neglecting fluctuations, the optimal velocity curve (OVC) of both
models can be derived analytically. This relation is helpful in order
to characterize the microscopic structure of the traffic phases
\cite{neu99,kno02,til00}.
\begin{figure}[!t] 
  \centerline{\epsfig{figure=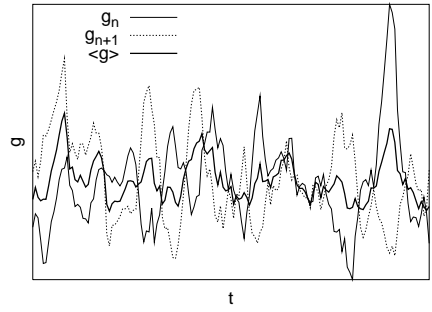,width=0.7\linewidth}}
  \caption{Time series of the gaps between two consecutive cars in the
    modeled chain of cars in a typical simulation.
    The leading cars drives at $V = 15
    ms^{-1}$. The mean gap $\langle g \rangle = (g_n + g_{n+1}) /2$
    varies hardly, since the two cars share a common
    gap $\propto 1/\rho$. The gaps of two consecutive cars are
    clearly anticorrelated.}
  \label{fig:osz}
\end{figure}

The OVC of the \SK\ results from its safety-condition
(\ref{eq:vsafe_sk}), i.e.
\begin{eqnarray}
  \label{eq:ovc1}
  \vs = -b\tau + \sqrt{b^2\tau^2 + \tilde{v}^2 + 2bg} = \tilde{v}.
\end{eqnarray}
\Eqref{eq:ovc1} is solved by $\tilde{v} = g / \tau$, therefore the OVC
of the \SK\ reads
\begin{eqnarray}
  \label{eq:ovc_sk}
  V_{\text{opt}}^{\text{SK}} (\rho) = \min \left\{ \frac{1}{\tau} \; \left( \frac{1}{\rho} 
  - \frac{1}{\rho_{\text{max}}}  \right), \vm \right\}.
\end{eqnarray}

From \eqref{eq:ovc_sk} it follows that $t_h^{\text{sk}} = \tau$ for the
deterministic case. The OVC asserts the results that time-headways
smaller than $\Delta t$ can not be modeled by the \SK\ since $\tau
\geq \Delta t$ is required due to safety constraints and the
stochasticity leads to a lowering of $\langle v \rangle$.

Regarding the \NSK\ two cases have to be distinguished (cf.
\eqref{eq:gc}). The OVC is derived from the condition
\begin{eqnarray}
  \label{eq:ovc2}
  \hat{v} &=& \vs =  \\
  \nonumber
  &-& b \tau 
  + \sqrt{b^2\tau^2 + \va^2 + 2b (g + \va \tau - \gamma_c)},
\end{eqnarray}
with
\begin{eqnarray}
  \label{eq:ovc3}
  \va = -b\tau + \sqrt{b^2\tau^2 + \hat{v}^2 + 2b\tilde{g}}.
\end{eqnarray}.
In the case of $\va > g_c$, i.e. $\gamma_c = g_c$, \eqref{eq:ovc2} is
solved by $\hat{v} \tau = (g + \tilde{g} - g_c)$ and the OVC reads
\begin{eqnarray}
  \label{eq:ovc_nsk_free}
  V_{\text{opt}}^{\text{f}} (\rho) = \min \left\{ \frac{2}{\tau} \; \left(
      \frac{1}{\rho} 
      - \frac{1}{\rho_{\text{max}}} - \frac{g_c}{2} \right), \vm
  \right\},\\
  \nonumber
  \qquad \va > g_c.
\end{eqnarray}
If $\va < g_c$, i.e., $\gamma_c =\va \, \tau$ the known expression of
the \SK\ follows,
\begin{eqnarray}
  \label{eq:ovc_nsk_jam}
  V_{\text{opt}}^{\text{j}} (\rho) = V_{\text{opt}}^{\text{SK}}
  , \qquad \va < g_c.
\end{eqnarray}
Therefore, in the high density regime, the \NSK\ behaves like the \SK\ 
as already stated with respect to the flow-density relation.
\subsection*{Follow--the--leader behaviour}
Finally, we investigate the differences in the follow--the--leader behaviour between
the two models. For this purpose we use a chain of 1000 cars
that follow the first car whose speed is fixed to
$V \le v_{\text{max}}$. The system is initialized by all cars
standing ($g_i = v_i = 0$). The zeroth car accelerates until the
constant velocity $V$ is reached.
Since then the system can be assumed to be stationary, 
quantities start to be measured once for the last car of the chain
$x \geq 10000 \cdot V$. Before
presenting the 
simulation results this set-up will be analysed more closely.
\begin{figure}[!t] 
  \centerline{\epsfig{figure=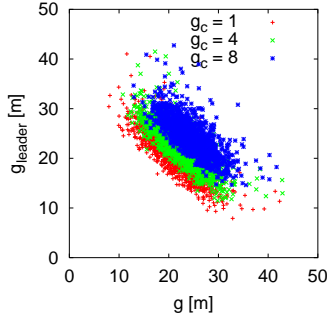,width=0.7\linewidth}}
  \caption{The simulations performed with the stochastic version of 
    the \NSK\ and for different values of $g_c$ are in qualitative
     agreement with \eqref{eq:antikor}.}
  \label{fig:gc}
\end{figure}

In the following just the deterministic case will be regarded. Then,
all the speeds can be eliminated to yield an update equation just for
the gaps. In order to keep the equations concise we adopt the
following notation: If a quantity is labeled with a prime $^\prime$ it
means timestep $t+\Delta t$, all others are to be taken at timestep
$t$. For the same reason $\tau = 1$ is used in the following.

Assuming a lead car driving at constant speed $v_0 = V$, the
behaviour of the \SK\ is then determined by the
equations
\begin{eqnarray}
  \nonumber
  v^\prime &=& v = - b + \sqrt{b^2 + V^2 + 2bg}\\
  \label{eq:fix_sk}
  g^\prime &=& g = g  + V - v^\prime
\end{eqnarray}
\Eqref{eq:fix_sk} has a fixed-point at $v^{\ast} = V$ and $g^{\ast} =
V$. Since this result can be expanded to the full chain of cars it
follows that
\begin{eqnarray}
  \label{eq:korr_sk}
  g_{n+1} = g_n.
\end{eqnarray}
The lower index denotes the position $n$ of the car in the chain.

The result explains the independence of the peak in the time--headway
distribution at low densities (cf. \figref{fig:th_free}). Moreover, it 
follows that the stochasticity of the \SK\ is not able to destroy the
fixed--point entirely. It is worth to say that the robustness of the
fixed--point in continuous car--following models is hard to overcome
and causes problems to model the synchronized state.

Now, the same situation is investigated for the \NSK. Again, the lead 
car drives constantly with $v_0 = V$ which is also $\va$ for the first
following car. Regarding the deterministic case, that car then
drives with $v_1^{\ast} = V$ and with constant headway $g_1^{\ast} =
g_c$. This is because
\begin{eqnarray}
  v_1^\prime = -b + \sqrt{(b + V)^2 + 2b(g_1-g_c)}
\end{eqnarray}
and
\begin{eqnarray}
  g_1^\prime = g_1 + V - v_1^\prime,
\end{eqnarray}
whose fixed--point $g_1^\prime = g_1 \equiv g_1^{\ast}$ is just
$g_c$. 

For the second car this procedure can be carried out to give
\begin{eqnarray}
  g_2^\prime = b + V + g_2 - \sqrt{b^2 + V^2 + 2bg_2},
\end{eqnarray}
where the stationary state of the first car $v_1^\ast = V$ and
$g_1^\ast = g_c$ has been assumed. This  equation has a simple
fixed--point, namely $g_2^{\ast} = V$. Obviously $v_2^{\ast} = V$ holds
alike. 

For the third car the computation leads to
\begin{eqnarray}
  \nonumber
  g_3^\prime  &=& b + V + g_3\\
  &-& \sqrt{b^2 - 2bg_c + V^2 + 2bg_2 + 2 b g_3},
\end{eqnarray}
and $g_3^{\ast} = g_c, v_3^{\ast} = V$. Generalized, the latter
equation reads
\begin{eqnarray}
  \nonumber
  g_{n+1}^\prime &=& b + V + g_{n+1}\\
  \label{eq:gen}
  &-& \sqrt{b^2 - 2bg_c + V^2 + 2bg_n + 2 b g_{n+1}},
\end{eqnarray}
leading to the following expression for the stationary state
$g_{n}^\prime = g_{n}$, $v_n = V$:
\begin{eqnarray}
  \label{eq:antikor}
  g_{n+1} = -g_n + g_c + V.
\end{eqnarray}
\begin{figure}[!t] 
  \centerline{\epsfig{figure=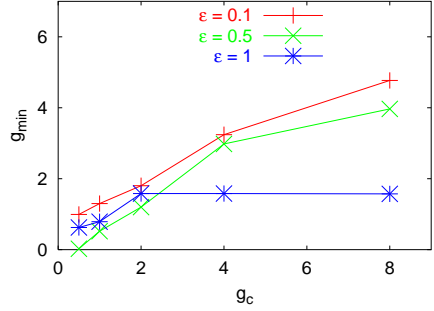,width=0.7\linewidth}}
  \caption{Minimal gap found in simulations of the loop taking
    different values for $g_c$ ($[m]$) and $\epsilon$.}
  \label{fig:safety}
\end{figure}
This result, \eqref{eq:antikor}, shows that asymptotically the gaps of 
the $n$--th 
and $(n+1)$--th car are anti-correlated. In \figref{fig:gc} time
series $g_n$ vs.\ $g_{n+1}$ are displayed for different values of
$g_c$. With increasing $g_c$ the corresponding line moves away from
the origin, therefore the corresponding flow decreases.

The result helps to understand the plateau structure found in the
time--headway distribution of the \NSK\ at low densities
(cf. \figref{fig:th_free}). With respect to \eqref{eq:antikor} the
time--headways of two consecutive cars cover the boun\-dary--points of
the interval $[g_c/V,1]$. Compared to the distribution of time
headways measured in simulation the lower bound will hardly be reached
due to stochasticity in the latter case.
As in the \SK\ stochasticity is not able to destroy
the fixed--point structure of the model, but two cars can exchange
their role in that structure, i.e., two cars share a gap given by mean
density but the share between $g_n$ and $g_{n+1}$ is not
fixed. \Figref{fig:osz} displays a time series of two cars sharing a
common gap.
\begin{figure}[!t] 
  \centerline{\epsfig{figure=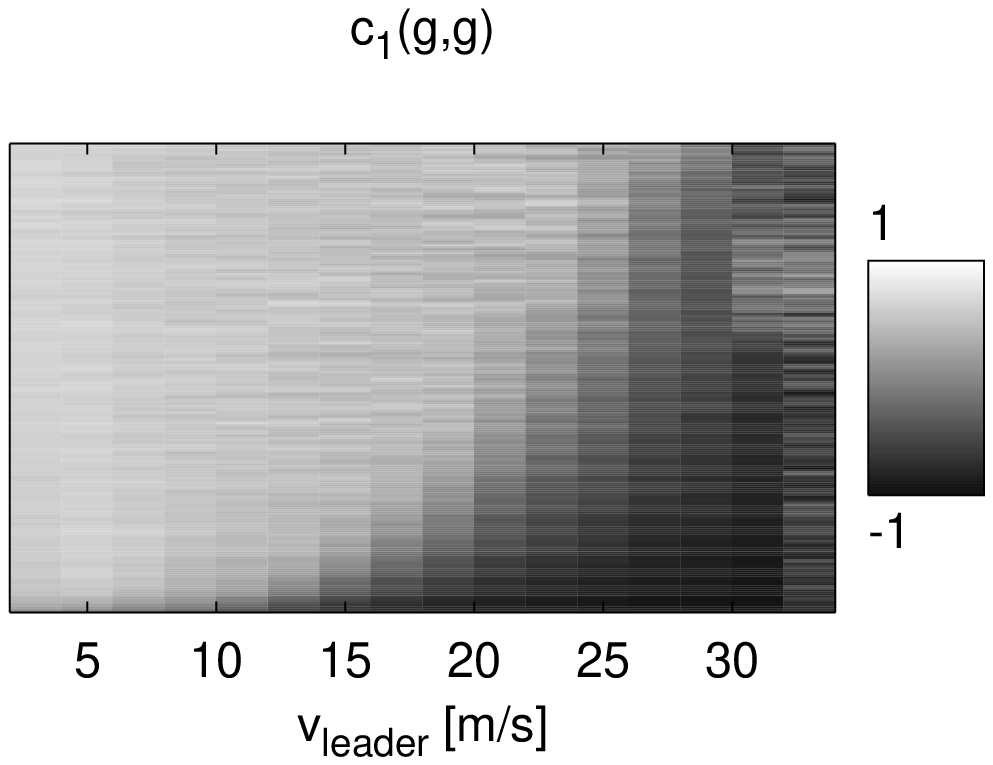,width=0.7\linewidth}}
  \centerline{\epsfig{figure=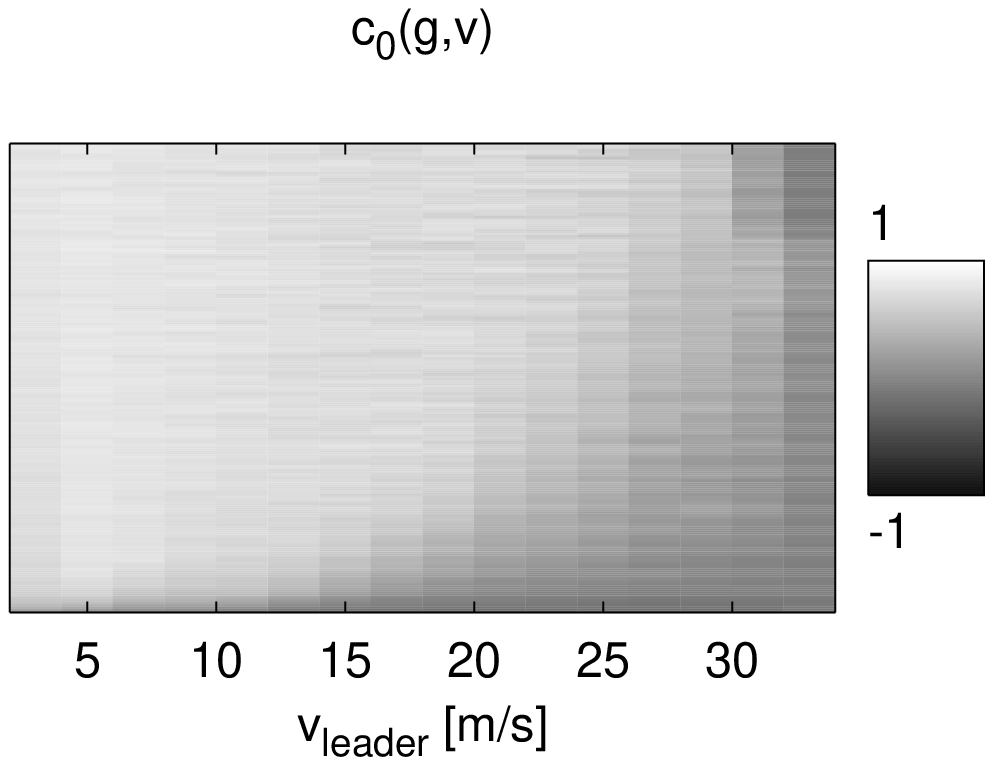,width=0.7\linewidth}}
  \caption{Correlation functions $c_1(g,g)$ (top)
    and $c_0(g,v)$ (bottom) for the follow---the--leader experimental
    setup in the \NSK. The correlations (value on the z-axis) have
    been calculated for each car in the chain, i.e. the value on the
    y-axis correspond to the car number $n$.}
  \label{fig:3dkorr}
\end{figure}

Before the correlation structure of the models is investigated in more
detail, the role of $g_c$ is shown. Looking at \eqref{eq:antikor}
one could wonder why the anticorrelation does not lead to states with
$g < 0$.

Assume that a car $(n+1)$ has closed in on its predecessor
$n$. Recalling \eqref{eq:gen}, the gap $g_{n+1}$ develops as
\begin{eqnarray}
  \nonumber
  g_{n+1}^\prime &=& b + V + g_{n+1}\\
  \nonumber
  &-& \sqrt{b^2 - 2bg_c + V^2 + 2bg_n + 2 b g_{n+1}}.
\end{eqnarray}
Setting  $g_{n+1} = g$ and $g_n \approx V$ the following approximation 
of $g_{n+1}^\prime$ holds,
\begin{eqnarray}
  \nonumber
  g_{n+1}^\prime &=&
  b + V + g - \sqrt{(b+V)^2(1 + \frac{2 b}{(V+b)^2}(g - g_c)} \\
  \nonumber
  &\approx& b + V + g - |b+V|  \left(1 - \frac{b}{(V+b)^2}(g -
    g_c)\right)
\end{eqnarray}
\begin{eqnarray}
  \label{eq:approx}
  = \left(1 -\frac{b}{V+b}\right) g + \frac{b}{V+b} g_c
\end{eqnarray}
\Eqref{eq:approx} shows, that once $g_n=V=const$, $g$ tends to
zero, but is finally stopped at $g_c$. A car that starts with $g<g_c$
is drawn towards $g_c$, which is the fixed-point. Therefore,
setting $g_c = 0$ safe driving can not be assured. This will get even
more clear if one determines the minimal gap that occurs over the full
range of densities dependent on the stochastic noise strength
$\epsilon$ and $g_c$, cf.~\figref{fig:safety}.
On the one hand, it can be seen clearly that the minimal gap found
increases with increasing $g_c$. On the other hand, the dependence on
$\epsilon$ is not so explicit. This results from the fact, that
$\epsilon$ does enter the model two--fold. Once, it acts similar as in
the \SK\ (cf. \eqref{eq:scheme_sk}) but it also determines $\va$.
With respect to the crash--free motion it can nevertheless been stated
that there is always a minimal $g_c^{\ast}$ which assures safe driving
if $g_c \geq g_c^{\ast}$ is chosen. But, the value of $g_c^{\ast}$
depends on the system parameters in a complicated way. It has to be
determined by simulation.

A closer look at the correlation function between two arbitrary
observables $\xi$ and $\chi$ at car $n$, $n+\Delta n$ respectively,
\begin{eqnarray}
  \label{eq:defkor}
  c_{\Delta n} (\xi_n,\chi_{n+\Delta n}) = \frac{\left\langle (\xi_n -
      \langle \xi_n \rangle) (\chi_{n+\Delta n} -  \langle
      \chi_{n+\Delta n} \rangle)\right\rangle}{\sigma_\xi^n
    \sigma_\chi^{n+\Delta n}}  
\end{eqnarray}
finally explains the effects of anticipation on the system's state. In
\eqref{eq:defkor} $\sigma_\xi^n$ stands for the standard deviation of
observable $\xi$ taken at car $n$.
\begin{figure}[!t] 
  \centerline{\epsfig{figure=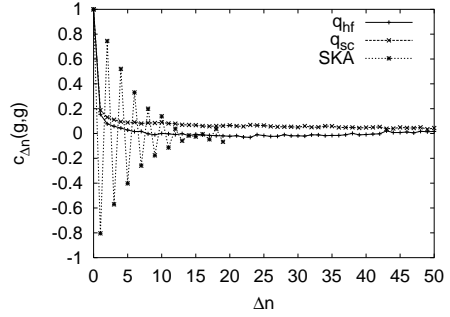,width=0.7\linewidth}}
  \caption{Correlation function $c_{\Delta n}(g,g)$ for single car
    data measured at the motorway junction Breitscheidt. Neither the
    traffic in high flow states $q_{\texttt{hf}}$ nor in synchronized
    states $q_{\texttt{sc}}$ show the strong anticorrelation found in
    the \NSK.}
  \label{fig:realcorr}
\end{figure}

In \figref{fig:3dkorr} the correlation functions $c_1(g,g)$ and
$c_0(g,v)$ are shown. In contrast to the \SK\footnote{%
In \SK\ $c_1(g,g) \approx 1$. The same holds for $c_0(g,v)$.}
there exist a platoon of cars behind the leading car which display a
strong anticorrelation between two consecutive gaps, cf.
\eqref{eq:antikor}. Note, that this structure is not destroyed if the
lead car also drives stochastically around $\langle v_{\text{leader}}
\rangle$. In the regime of strong anticorrelation $c_0(g,v) \approx
0$ a car is somehow free to choose a gap $g$ independent of the speed
of the leading car. Therefore, this state reminds one of synchronized
flow\cite{ker98,kera,kerb,kerc}, a traffic state that mostly occurs at
bottlenecks.  However, further exploration of the simulated data show
that the velocity only displays fluctuations of strength $a/2$ around
its mean speed $\langle v_{\text{leader}} \rangle$, therefore
anticipation alone is not able to generate a synchronized
state. Moreover, the strong anticorrelation found in the \NSK\ 
is not present in measured data, cf. \figref{fig:realcorr}.
\section{Conlusions}
The effects of anticipatory driving are investigated by means of
simulation as well as analytical calculations. As a reference for
comparison a well--understood traffic flow model
\cite{kra97,kra98,kra98b} was enhanced by next--nearest--neighbour
interaction and investigated under certain aspects. Simulation results 
show that the introduction of anticipation does not lead to changes in 
the mechanism that generates wide moving jams nor does the 
dynamics inside a jam changes. However, one observes a stabilization
of the flow in dense traffic, which is "crucial to overcome the
difficulties in describing the empirically observed phases and their
transitions" \cite{kno02b}. Moreover, as one could have been
expecting, the ``optimized'' driving strategy leads to very short
temporal headways under free flow conditions. Such short headways are
also found in measurements on streets. Their existence can be
explained quite general by exploring the mechanism that is introduced
in the model by next--nearest--neighbour interaction. 

In the present article it is shown
that this general mechanism works by
coupling three cars 
together to share their two respective headways: if one car is fairly 
close to the car in front, then its follower has to hold a distance
that is roughly equal to the average speed of the car-ensemble.
Obviously, this can be generalized to many cars in front, leading in
principle to a situation that has been envisioned already by the
automobile industry: platoons of cars that are electronically coupled
to optimize the energy consumption by driving with very small headways
at very large speeds. There, this goal is achieved by very small
reaction times of the control system, while an approach based on the
ansatz chosen for car-following with anticipation just needs
communication to more than one car ahead, but more human-like reaction
times. 

Surprisingly, despite the fact that the approach of sharing headways
sounds fairly natural, it is not easy to show that it happens in
reality also. Single-car data (\figref{fig:realcorr}) show no sign of
anticorrelation in car headways, whereas all models working with
anticipation display clearly. Beside the \NSK\ we investigated the
BL--CA \cite{kno00} as a representative for a cellular--automaton
approach in the same way in order to confirm that assertion.
Therefore, anticipation alone is not able to explain all traffic
states and their microscopic behaviour found in observations. 

The authors like to thank Andreas Schadschneider for useful
discussions. 

\end{document}